# Detecting extreme event-driven causality

Siyang Yu[1], Yu Huang[2,3], Zuntao Fu[1*]

1. Laboratory for Climate and Ocean-Atmosphere Studies, Department of Atmospheric and Oceanic Sciences, School of Physics, Peking University, Beijing, China
2. Earth System Modelling, School of Engineering and Design, Technical University of Munich, Munich, Germany.
3. Complexity Science, Potsdam Institute for Climate Impact Research, Potsdam, Germany.

*Correspondence to: fuzt@pku.edu.cn

**Abstract**：

The occurrence of some extreme events (such as marine heatwaves or exceptional circulations) can cause other extreme events (such as heatwave, drought and flood). These concurrent extreme events have a great impact on environment and human health. However, how to detect and quantify the causes and impacts of these extreme events by a data-driven way is still unsolved. In this study, the dynamic system method is extended to develop a method for detecting the causality between extreme events. Taking the coupled Lorenz-Lorenz systems with extreme event-driven coupling as an example, it is demonstrated that this proposed detecting method is able to capture the extreme event-driven causality, with even better causality detecting performance between concurrent extreme events. Comparison among three kinds of measured series, full measurements outperform partial ones in event-to-event causality detecting. The successful applicability of our proposed approach in Walker circulation phenomenon indicates that our method contributes a novel way to the study of causal inference in complex systems. This method offers valuable insights into multi-scale, nonlinear dynamics, particularly in uncovering associations among extreme events.

**Keywords**： Extreme event, causality, dynamic system, cause and effect, event-to-event

# 1. Introduction

As global warming increases, extraordinary extreme events, such as heatwaves or floods, occur frequently [1-3]. Especially, some extreme events occur simultaneously across different regions [4-9], such as co-occurrence of Pakistan flooding and East Asian heatwaves during July to August 2010 and 2022 [7, 8]. These co-occurring extreme events lead to more serious casualties and property losses [7-10]. Robust in-phase relationship between the two extreme events has been reported in several studies [4-9], for example, severe droughts in South America have been found to coincide with marine heatwaves in the South Atlantic, which may be caused commonly by convection over Indian Ocean stirring by MJO [4]. Much effort has also been devoted to studying the physical drivers behind these extreme events [11-18], and persistent blocking and Rossby wave trains have been reported to play an important role in forming these extreme events. However, correlations or links don't indicate causality, two in-phase extreme events may have different cause-effect relation. Furthermore, not each persistent blocking can induce its corresponding extreme events or not all extreme events are induced from blocking [4-6, 11-18]. The causes of each extreme event may have individually different sources. Moreover, the issue of data-driven causality detection between the two specific extreme events remains unsolved [17-20].

Traditionally, the causes of observed phenomena involving reported extreme events are uncovered through temporally evolutionary processes in which studied events occur over a certain time interval or at a specific moment. Several approaches based on this perspective have been developed to detect causal relations between the underlying processes, such as Granger causality (GC thereafter) and its extended variants [21-24], transfer entropy (TE thereafter) [25], convergent cross mapping (CCM thereafter) [26-30], machine learning based methods [31, 32] and so on. However, all these approaches [21-36] are limited to detecting global or local average causality, and are unable to capture event-to-event causality, even though the states of system [37] or event-related time-dependent directional couplings [38] have been taken into accounts.

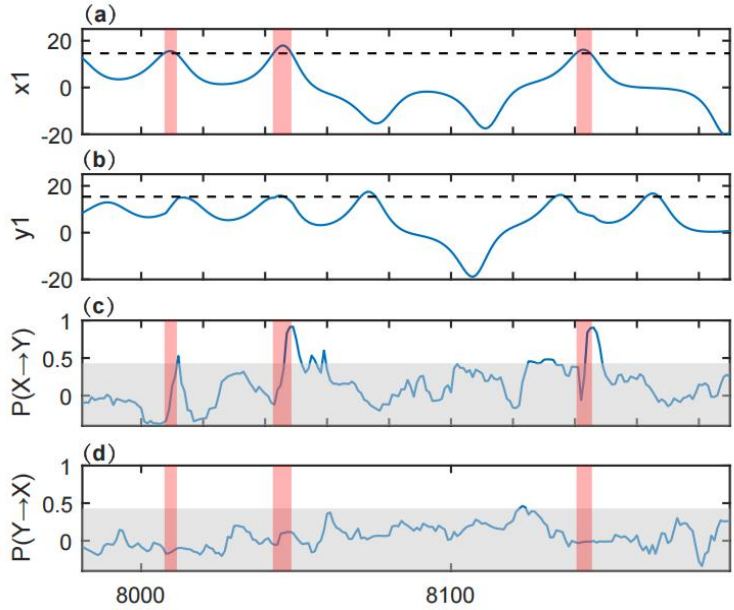

**Fig. 1** Causality detecting from full measurements in the coupled Lorenz-Lorenz model with a unidirectional extreme event-driven coupling from **X** to **Y**. Typical series of partial measurements x1 (a) and y1 (b) for the two systems **X** and **Y**. Together with the calculated instantaneous causality index for (c) from **X** to **Y** and (d) from **Y** to **X**, respectively. The red shadings represent the periods when x1 exceeds its respective 95th percentile values, and the gray shading represents the 98% significance interval of the instantaneous causality index.

The average causality detection, whether global or local, may fail to capture the true causality in an extreme event-driven system. Taking a coupled Lorenz-Lorenz model with unidirectional extreme event-driven coupling (see subsection **2.1** for details) as an example (**Fig. 1**), as expected, the Granger-based causality detecting methods cannot capture the ground-truth cause-effect relation from system **X** to **Y** (Figure not shown here), since Granger-based methods fail in nonlinear processes or systems [18, 26, 27, 31-34]. Similarly, classical TE (**Fig. 2b**) [25], Peter-Clark Momentary Conditional Independence (PCMCI thereafter) [39, 40] and Information Imbalance Causality (IIC thereafter) [41] also suffer from this similar limitation (Figure not shown here). Moreover all these methods are designed to reveal the average causality [37]. The CCM is designed to capture the average cause-effect relation in nonlinear dynamical systems, which can indeed detect a certain percentage of the average true causality at low confidence level. However, as the confidence level increases, only a much smaller proportion of the average ground-truth causality is detected (**Fig. 2a**), making the detection of event-to-event ground-truth causality even more challenging.

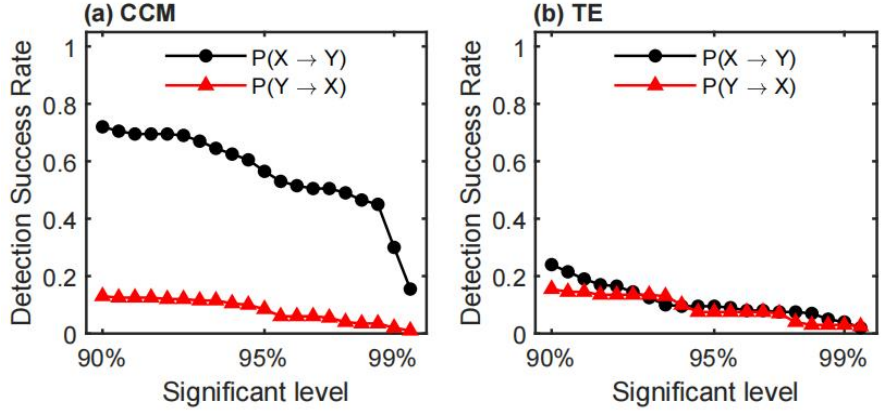

**Fig. 2** Causality detecting from partial measurements (only x1 and y1) in the coupled Lorenz-Lorenz model with a unidirectional extreme event-driven coupling from **X** to **Y** by (a) CCM, (b) TE. Here, the detection success rate is defined based on 100 samples, each containing 10000 data points, and reflects the number of samples in which the global average causality is correctly identified.

In the real-world scenarios, interactions between variables in complex systems are usually relate to the interactions among extremes. For example, tropical sea surface temperature strongly influences other climate variables only during the extreme conditions, El Niño or La Niña events. However, despite both being active areas of research, few studies have directly investigated the causality of extreme events [18-20]. Zanin [19] and Gnecco et al [20] examined causality of extremes by means of a probability related way, which is still an average causality detection method to answer the question "do extremes in one variable cause extremes in another variable?". While Rényi conditional mutual information (RCMI thereafter) designed by Paluš and his colleagues [18] answers the question "which variable is the likely cause of extremes in the affected variable?". However, all these average causality detection methods cannot answer the following question "is a specific extreme event in one variable a cause or a response to a corresponding extreme event in another variable?". To address this type of problems, an event-to-event causality detecting (E2ECD thereafter) method is required to develop to be able to detect causality between extreme events. This is the objective of this study.

The rest of this study is organized as follows. Section 2 describes a modified Lorenz-Lorenz coupled model with extreme event-driven couplings and the data sets related to Walker circulation used for the subsequent analysis. Section 3 presents a framework and its detailed steps related to the proposed E2ECD method. To demonstrate the performance of our E2ECD method, Section 4 is divided into two subsections. One is the performance of event-to-event method in causality detection for extreme events in the Lorenz-Lorenz coupled model with extreme event-driven couplings, the other is the application of event-to-event method in causality detecting for extreme events in Walker circulation in climate system. Section 5 concludes this study with some discussions about future work.

## 2. Model and data description

### 2.1 Coupled Lorenz-Lorenz model with extreme event-driven coupling

The coupled Lorenz-Lorenz model about $\mathbf{X} = \{x_1(t), x_2(t), x_3(t)\}$ and $\mathbf{Y} = \{y_1(t), y_2(t), y_3(t)\}$ is defined from the classic Lorenz model [42] and it can be written as follows [38, 43]:

$$\mathbf{X} \begin{cases} \dot{x}_1 = p(x_2 - x_1) + \nu y_1 \\ \dot{x}_2 = r x_1 - x_2 - x_1 x_3 \\ \dot{x}_3 = x_1 x_2 - b x_3 \end{cases} \tag{1}$$

$$\mathbf{Y} \begin{cases} \dot{y}_1 = p'(y_2 - y_1) + \alpha x_1 \\ \dot{y}_2 = r' y_1 - y_2 - y_1 y_3 \\ \dot{y}_3 = y_1 y_2 - b' y_3 \end{cases} \tag{2}$$

with $p = 10$, $r = 35$, $b = 8/3$, $p' = 10$, $r' = 39$, $b' = 8/3$. And $\nu$ and $\alpha$ are coupling parameters with controllable values, from which the cause-effect relation and direction can be presented. This coupled model can serve as a benchmark to evaluate the performance of causality detection methods, including our proposed method, showing their advantages and limitations.

The 4$^{th}$ order Runge-Kutta routine with integration step size and sampling interval of both 0.02 was used to generate 100,000 data points firstly with a set of initial values and the coupling parameters set as $\nu = \alpha = 0$. From the partial measurements $x_1$ and $y_1$, some specific thresholds (such as the 95th percentile used in this study) can be calculated. Based on the same set of initial values and determined thresholds, an ensemble of 10 samples, each with 10,000 points can be generated under different coupling settings.

1) Unidirectional coupling with the extreme event driving

Set $\nu = 0$ and when $x_1$ exceeds its 95th percentile, set $\alpha = 4$, integrating the coupled system to generate an ensemble. For each sample, there is a unidirectional driving from **X** to **Y**.

2) Bidirectional coupling with the extreme event driving

When $x_1$ exceeds its 95th percentile, set $\alpha = 4$; when $y_1$ exceeds its 95th percentile, set $\nu = 4$, integrating the coupled system to generate an ensemble. For each sample, there is a bidirectional driving from **X** to **Y** and from **Y** to **X**.

3) Intermittent coupling with the extreme event driving

For the first 3000 points, set $\nu = 0$, when $x_1$ exceeds its 95th percentile, set $\alpha = 4$; for the subsequent 4,000 data points, set $\nu = \alpha = 0$; for the last 3000 points, set $\alpha = 0$, when $y_1$ exceeds its 95th percentile, set $\nu = 4$. Integrating the coupled system to generate an ensemble. For each sample, there is an intermittent coupling with the first 3000-point **X** driving **Y** and last 3000-point from **Y** driving **X** linked with 4000-point no coupling.

## 2.2 Data related to Walker circulation

The monthly data from the Modern-Era Retrospective analysis for Research and Applications Version 2 (MERRA-2) [44] are employed to assess the causal relationship between the western and eastern Pacific subsystems within the Walker circulation, for the period ranging from 1980 to 2021. Variables include the sea level pressure (SLP) and 2-meter air temperature (T2m) and are analyzed at a 2.5° longitude by 2° latitude resolution. The regional averaged SLP over the region 120°E–180°, 0°–10°N is used as the variable series for the western Pacific (WPAC), while regional averaged T2m over the region 160°W – 130°W, 0° – 10°N is used as the variable series for the central and eastern Pacific (CEPAC) [37].

## 3. Instantaneous Causality detecting Method

To detect the causality between extreme events in two systems or processes, an event-to-event instantaneous causality detecting method is developed and summarized as the following steps:

**(1)** Import time series of the two systems, one is $\boldsymbol{X}(t) = (x_1(t), x_2(t), \ldots, x_N(t))$ and the other is $\boldsymbol{Y}(t) = (y_1(t), y_2(t), \ldots, y_{N'}(t))$, where $t = 1, 2, \ldots, T$, and T is the data length. For the coupled Lorenz-Lorenz model used in this study, T=10000.

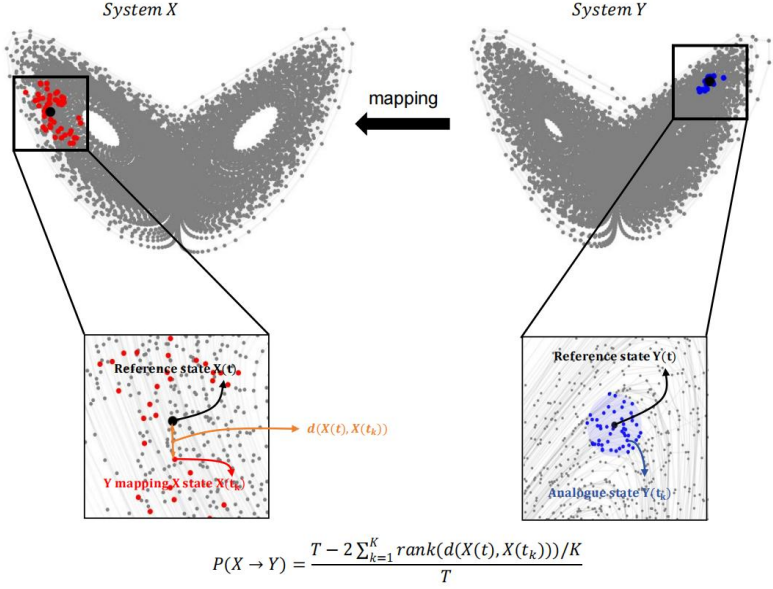

**Fig. 3** Illustration of the instantaneous causality detection method, taking the causality from X to Y as an example. Dark black points are reference points and blue (red) points are their corresponding neighbors (mapping conditioned-neighbors).

**(2)** The distances between different time points within each system can be quantified by calculating their Euclidean distances $dX(t, t_k)$ or $dY(t, t_k)$:

$$dX(t, t_k) = \|X(t) - X(t_k)\| =$$
$$\sqrt{(x_1(t) - x_1(t_k))^2 + (x_2(t) - x_2(t_k))^2 + \ldots + (x_N(t) - x_N(t_k))^2} \quad (3)$$

$$dY(t, t_k) = \|Y(t) - Y(t_k)\| =$$
$$\sqrt{(y_1(t) - y_1(t_k))^2 + (y(t) - y_2(t_k))^2 + \ldots + (y_{N'}(t) - y_{N'}(t_k))^2} \quad (4)$$

For a reference state $Y(t_k)$ (see the bold black point shown in **Fig. 3**), we identify the K nearest neighbors and their corresponding time index $t_k$ ($0 < k \leq K$), in this study, set $K = 10$. Here $N$ and $N'$ are the number of variables in system $X(t)$ and $Y(t)$, respectively. For coupled Lorenz-Lorenz model used in this study, $N = N' = 3$.

**(3)** Map $Y(t_k)$ into state space of **X** to obtain $X(t_k)$, i.e., **Y**-conditioned states in **X**. Calculate the Euclidean distance between $\mathbf{X}(t_k)$ and $\mathbf{X}(t)$ for each neighbor, and then perform an ascending order sorting among all states in **X** to obtain its ranking index $GX(t, t_k) = rank(dX(t, t_k))$. $(GX(t,t) = 0, GX(t,t_k) \leq T - 1)$.

**(4)** After normalization, the causality index is obtained as:

$$P(X \to Y) = \frac{T - 2\sum_{k=1}^{K} rank(d(X(t), X(t_k)))/K}{T}, t \in [1, T], P(X \to Y) \in (-1, 1)$$

(5)

Similarly, the causality index from opposite direction can be defined as

$$P(Y \to X) = \frac{T - 2\sum_{k=1}^{K} rank(d(Y(t), Y(t_k)))/K}{T}, t \in [1, T], P(Y \to X) \in (-1, 1)$$

(6)

**(5)** A delay time $\tau$ is introduced to address the situation time-delayed causal influence within two systems, i.e., input $\mathbf{X}(t) = (x_1(t), x_2(t), \ldots, x_N(t))$, $\mathbf{Y}(t + \tau) = (y_1(t + \tau), y_2(t + \tau), \ldots, y_{N'}(t + \tau))$ and calculate the causality index $P_{X \to Y}(t, \tau)$ by repeating steps from (2) to (4).

**(6)** Statistical significance test: first generate the surrogate data by shuffling the temporal order of any variable (For example, $\mathbf{X} \to \mathbf{X}_s$), and then calculate the causal index $P(X_s \to Y)$ by repeating steps from (1) to (4). Repeat this calculation for a number of times (such as 1000) to form a probability distribution function (PDF) of the causal index where the coupling is absent. The causality index $P(X \to Y)$ is significant at the 95% or even higher confidence level, when it exceeds the 95th or even higher percentile of the causal index $P(X_s \to Y)$. The basic criterion behind this causality detecting is that if **X** is causally influencing **Y**, the information of **X** is included into the dynamics of **Y**, and thus two close states on the phase space of **Y** correspond to two close states on the phase space of **X** (**Fig. 4a**);

inversely, if there is no cause-effect relation between **X** and **Y**, the information of **X** is not included in the dynamics of **Y**, and thus two close states on the phase space of **Y** do not correspond to two close states on the phase space of **X** (**Fig. 4b**).

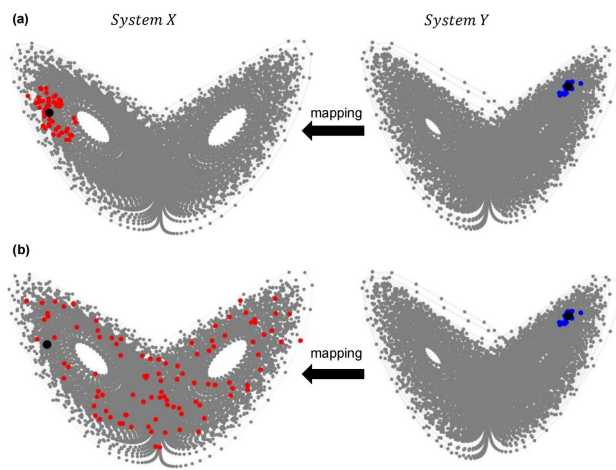

**Fig. 4** Schematic presentation of phase space points under the condition (a) **X** causes **Y**, (b) **X** does not cause **Y**. Dark black points are reference points and blue (red) points are their corresponding neighbors (mapping conditioned-neighbors).

(7) Identify the causal direction: taking **X** causing **Y** as an example, if $\max_\tau(P(X \to Y))$ is statistically significant and meanwhile with $\tau_{max} \geq 0$, then it is considered that there is a causal relationship from **X** to **Y** at time t (as shown in **Fig. 5**). Here, the condition of $\tau_{max} \geq 0$ is due to the fact that only the past information of **X** can affect the future information of **Y**.

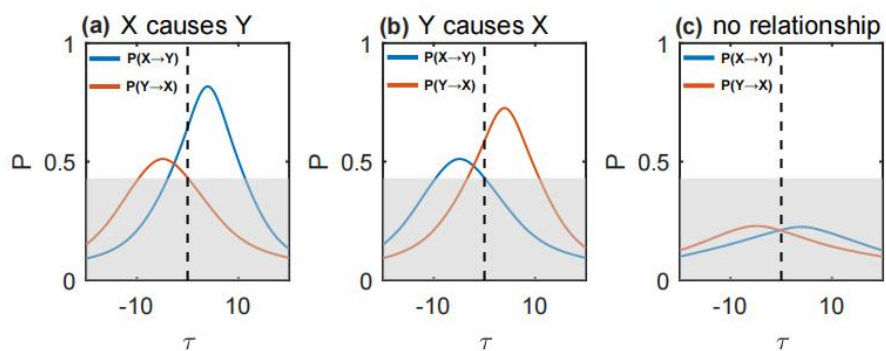

**Fig. 5** Schematic for identifying causality and its direction. (a) **X** causes **Y**, (b) **Y** causes **X**, (c)

no cause-effect relation between **X** and **Y**. Grey shading represents the 98% significance interval for the surrogate instantaneous causality index.

## 4. Results
## 4.1 Event-to-event causality detection in coupled Lorenz-Lorenz system with extreme event-driven coupling
### 4.1.1 Event-to-event causality detection

First of all, each extreme event in system **X** doesn't necessarily drive its own response extreme event in system **Y** (**Fig. 1a and 1b**). For the unidirectional extreme event-driven cases, only 23% extreme events in driving system **X** co-occur with extreme events in driven system **Y**. This indicates that not all extreme event drivers can lead to their response extreme events generated, which may influence the efficiency of event-to-event causality detection.

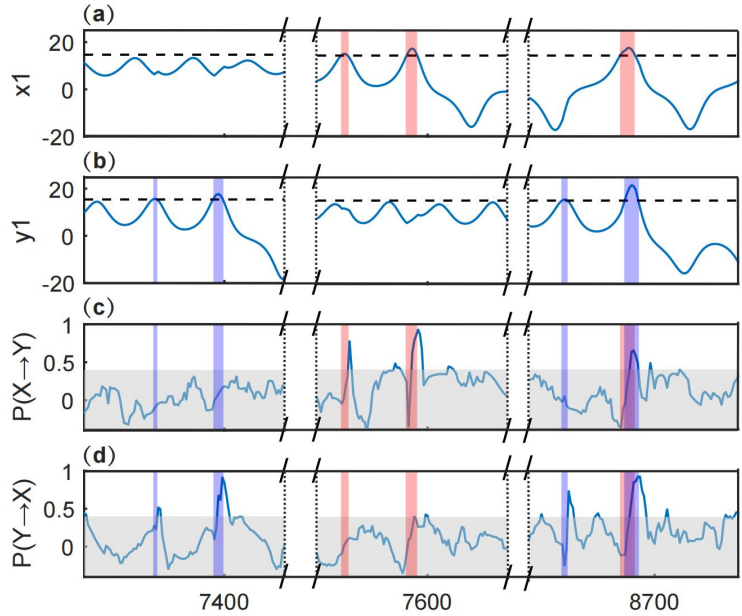

**Fig. 6** Causality detection from full measurements in the coupled Lorenz-Lorenz model with a bidirectional extreme event-driven coupling. Typical series of partial measurements x1 (a) and y1 (b) for the two systems **X** and **Y**, showing **Y** drives **X** (left), **X** drives **Y** (middle) and bidirectional coupling between **X** and **Y** (right), respectively. Together with the calculated instantaneous causality index for(c) from **X** to **Y** and (d) from **Y** to **X**, respectively. The red and blue shading represent the time periods when x1 and y1 exceed their respective 95th percentile values, and the gray shading represents the 98% significance interval of the surrogate instantaneous causality index.

The E2ECD method can effectively capture the event-to-event causality preset in the coupled Lorenz-Lorenz system with both unidirectional (**Fig. 1**) and bidirectional (**Fig. 6**) extreme event-driven couplings. Near the position of each extreme event highlighted by color shadings, the event-to-event causality and direction can be correctly detected (**Fig. 1 and Fig. 6**). It should also be pointed out that the position determined by E2ECD method are not always totally overlapped with the position of each extreme event driver, but with a little bit delay (**Fig. 1c, Fig. 6c and 6d**), which might be attributed to the nonlinear interactions in the coupled Lorenz-Lorenz system. This also suggests that the efficiency and accuracy of E2ECD method can be further enhanced by optimizing the settings of the hyper-parameters, which will be demonstrated in the next subsection.

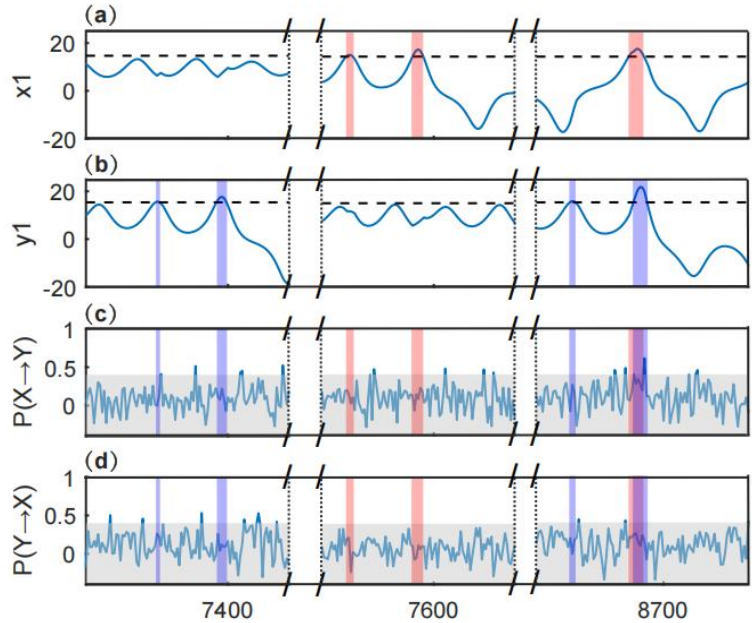

**Fig. 7** Same as **Fig. 6** but from partial measurements (only x1 and y1) in the coupled Lorenz-Lorenz model with a bidirectional extreme event-driven coupling.

In the real-world systems, full measurements are typically unattainable, only partial measurements of subsystems or projections of true systems can be obtained. To examine the causality detection performance in the case of partial measurements, here we suppose that only $x_1(t)$ (denoted as x1) in system **X** and $y_1(t)$ (denoted as y1) in system **Y** are available. Combined with the E2ECD method, causality between systems X and Y can be inferred in two strategies using partial measurements: one is directly from both x1 and y1 and denoting this way as P2 (**Fig. 7**), the other from both x1 and y1 together with state space reconstruction (embedding parameters: embedding dimension set as m=10 and time delay as lag=6), denoting this way as P3 (**Fig. 8**). Compared with the results from full measurements (denoting as P1 **Fig. 6**), E2ECD can still identify the ground-truth event-to-event causality, though with reduction in efficiency and accuracy by P2 and P3.

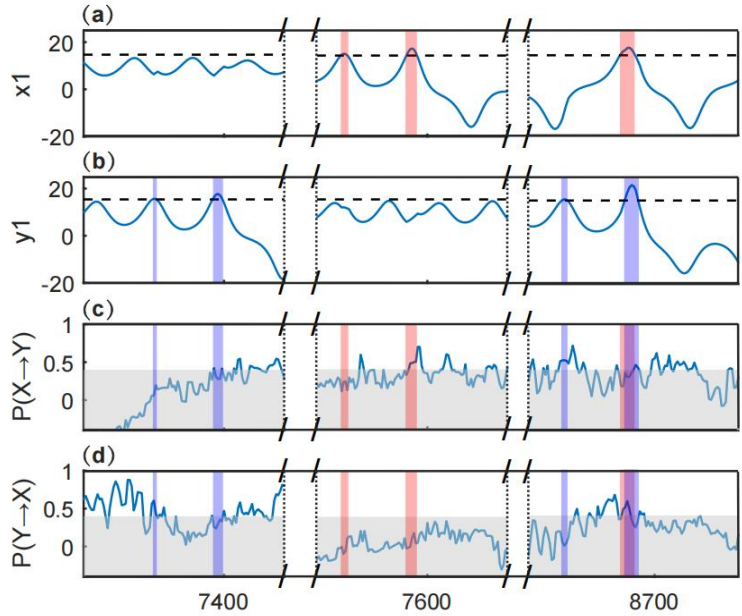

**Fig. 8** Same as **Fig. 6** but from partial measurements (only x1 and y1 with state space reconstruction) in the coupled Lorenz-Lorenz model with a bidirectional extreme event-driven coupling.

More details for comparison among P1, P2 and P3 together with comparison between unidirectional and bidirectional coupling under different significance levels are summarized in **Fig. 9**. To quantify the efficiency and accuracy of event-to-event causality detection under different significance levels, we use a metric that measures the fraction of correctly identified extreme event-driven events by E2ECD relative to the total number of such events, within a specified detection window of $dt = 11$.

Firstly, as the significance level increases, the efficiency and accuracy of all three ways (P1, P2 and P3) decrease as expected, unidirectional and bidirectional coupling. Secondly, among the three ways, overall P1 performs best, and P3 better than P2, especially at higher significance level. Thirdly, the three strategies perform better in the bidirectional coupling scenario than in the unidirectional coupling scenario, which may be caused by more extreme events taking part in forming cause-effect relation in the bidirectional scenario.

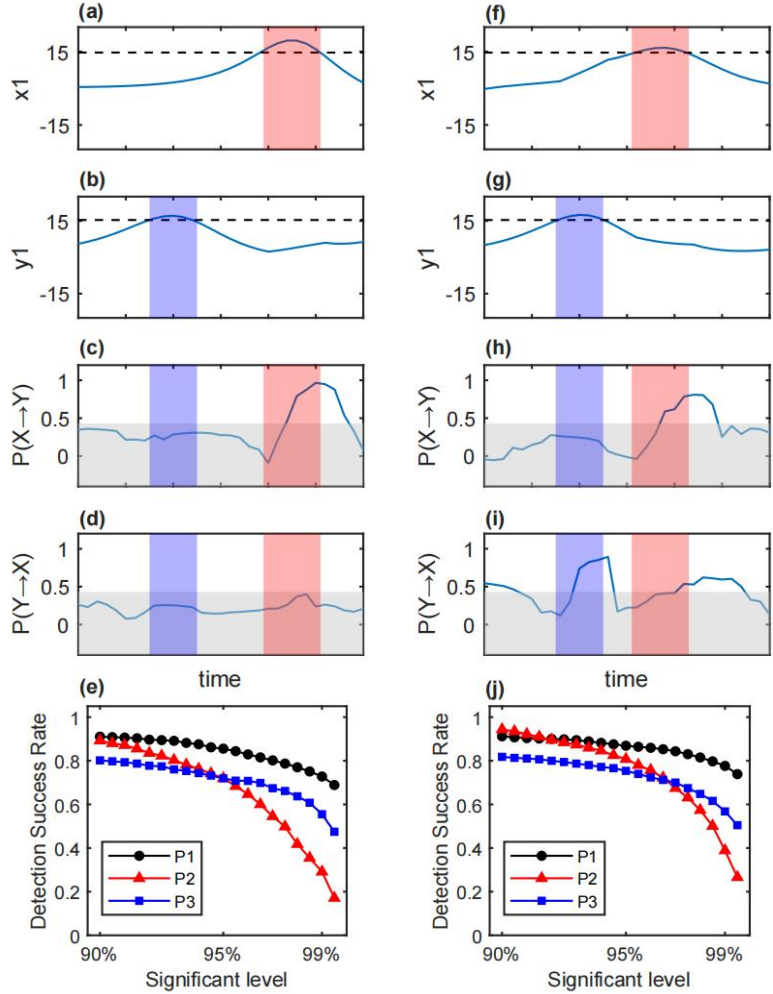

**Fig. 9** Causality detection in the coupled Lorenz-Lorenz model under the unidirectional coupling scenario (a, b, c, d, e) and under the bidirectional coupling scenario (f, g, h, i, j). (e) and (j) show the fraction of correctly identified extreme event-driven events by causality index to total extreme event-driven events, across different significance levels, under the unidirectional coupling scenario (with a detection window of $dt = 11$). P1 for full measurements, P2 for partial

measurements (only x1 and y1), and P3 for partial measurements (only x1 and y1 with state space reconstruction).

In more complex scenarios involving intermittent extreme event-driven coupling, including alternate drives from X to Y, periods without coupling, and Y driving X, the E2ECD method still performs effectively (**Fig. 10**). Since the E2ECD method is specifically designed for event-to-event causality detection, it is well-suited for handling time-varying causality due to its strong capability for detecting instantaneous causal relationships.

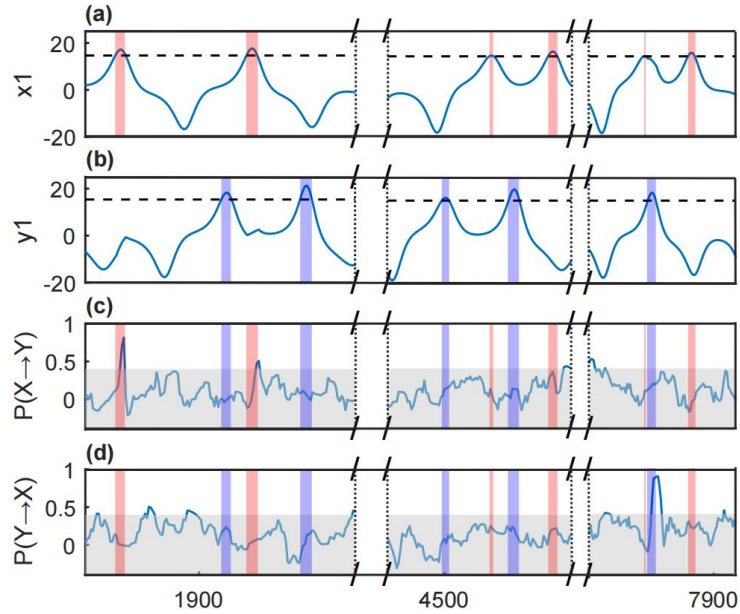

**Fig. 10** Causality detection from full measurements in the coupled Lorenz-Lorenz model with an intermittent extreme event-driven coupling. Typical series of partial measurements x1 (a) and y1 (b) for the two systems **X** and **Y**, showing **X** drives Y (left), no coupling between **X** and **Y** (middle) and **Y** drives X (right), respectively. Together with the calculated instantaneous causality index for (c) from **X** to **Y** and (d) from **Y** to **X**, respectively. The red and blue shading represent the time periods when x1 and y1 exceed their respective 95th percentile values, and the gray shading represents the 98% significance interval of the instantaneous causality index.

### 4.1.2 Robustness test for hyper-parameters and experiment settings

As shown in **Fig. 6**, the position determined by E2ECD method are not perfectly overlapped with the position of each extreme event driver, but with a tiny time delay (**Fig. 1c, Fig. 6c and 6d**). Here we demonstrate that the imperfect performance of the E2ECD method can be explained by the hyper-parameters and experiment settings, such as detection window width used to determine whether there exists cause-effect relationship between two systems or processes, the duration or intensity of extreme event drivers, and co-occurrence of extreme events in both driving and driven variables.

First of all, two detection window widths $dt = 11$ and $dt = 5$ are used to show the impacts of detection window width on determining whether there exists cause-effect relationship between the two systems (**Fig. 11**). The efficiency and accuracy of event-to-event causality detection are quantified by the detection success rate, which counts the number of extreme event drivers correctly identified by the E2ECD method relative to the total number of such drivers. Overall, the results show that this performance is not highly sensitive to the width of the detection window under different significance levels. The three strategies (P1, P2 and P3) exhibit similar behavior in the robustness tests. Narrowing detection window width reduces the efficiency and accuracy of event-to-event causality detecting, but not to a degree that qualitatively changes the overall conclusions.

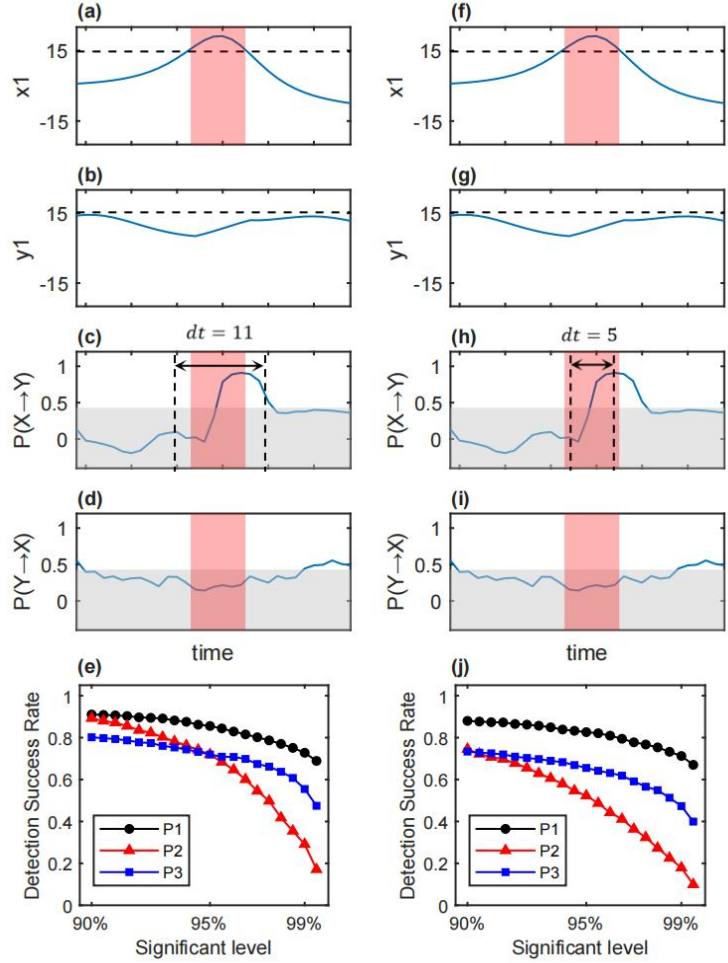

**Fig. 11** Robustness test on detection window width in causality detection in the coupled Lorenz-Lorenz model with a unidirectional extreme event-driven coupling. (a), (b), (c), (d), (e) for detection window width $dt = 11$, and (f), (g), (h), (i), (j) for detection window width $dt = 5$.

Secondly, 66% of the driving extreme events last for at least 5 time points, while only 34% persist for less than 5 points. This indicates that most extreme events tend to have longer durations. The varying duration of these driving extreme events can significantly influence the detection of event-to-event causality. As expected, the E2ECD method demonstrates reliable performance when applied to full measurements (**Fig. 12**). However, the results given by the E2ECD method from partial measurements (both P2 and P3) are nearly insensitive to the duration of driving extreme events (**Fig. 12**). This may be due to the loss of duration information from full measurements to partial measurements. Previous studies have found that global warming drives a threefold increase in persistence and 1°C rise in intensity of marine heatwaves [45], these findings together with the results shown in **Fig. 12** indicate that if we can obtain the full measurements about the marine heatwaves, the increase of marine heatwaves' duration can help us easily uncover their causal impacts on other marine ecosystems.

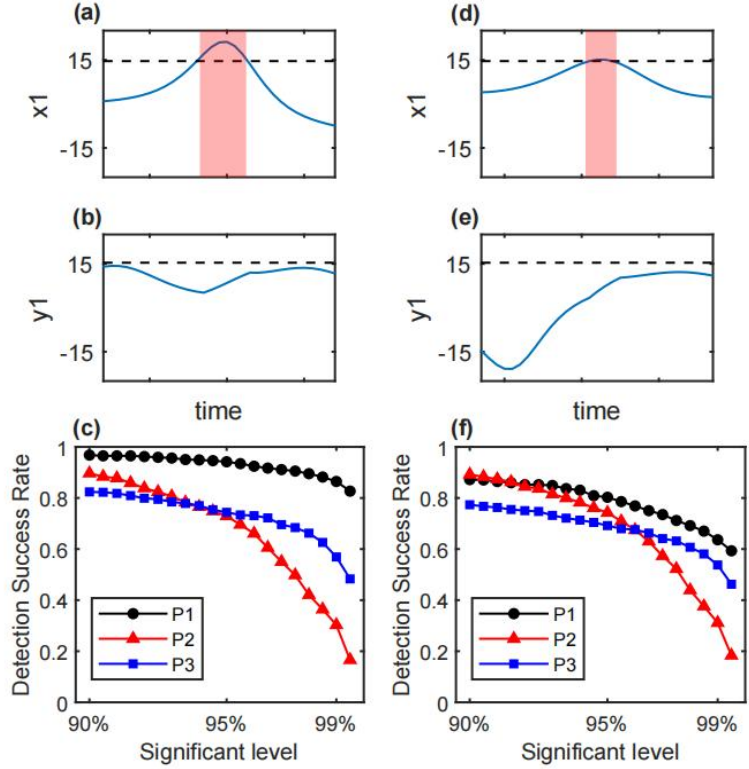

**Fig. 12** Robustness test on duration of extreme events in causality detecting with detection window width $dt = 11$ in the coupled Lorenz-Lorenz model with a unidirectional extreme event-driven coupling. (a), (b), (c) extreme events with a duration not less than 5 points, and (d), (e), (f) extreme events with a duration of less than 5 points.

The intensity of extreme events is closely related to their duration, i.e. stronger extreme events last longer. Accordingly, we conjecture that the impacts of intensity of extreme events on detecting of event-to-event causality by E2ECD may be similar to those from the duration. If we define high-intensity extreme events as those exceeding the 98th percentile and low-intensity extreme events as those with a maximum value exceeding the 95th percentile but below the 98th percentile, we find that there 44% driving extreme events are of high-intensity and 56% of low-intensity. A comparison of event-to-event causality detection by E2ECD between high-intensity and low-intensity extreme events reveals similar behavior to that observed for event duration, as expected (**Fig. 13**).

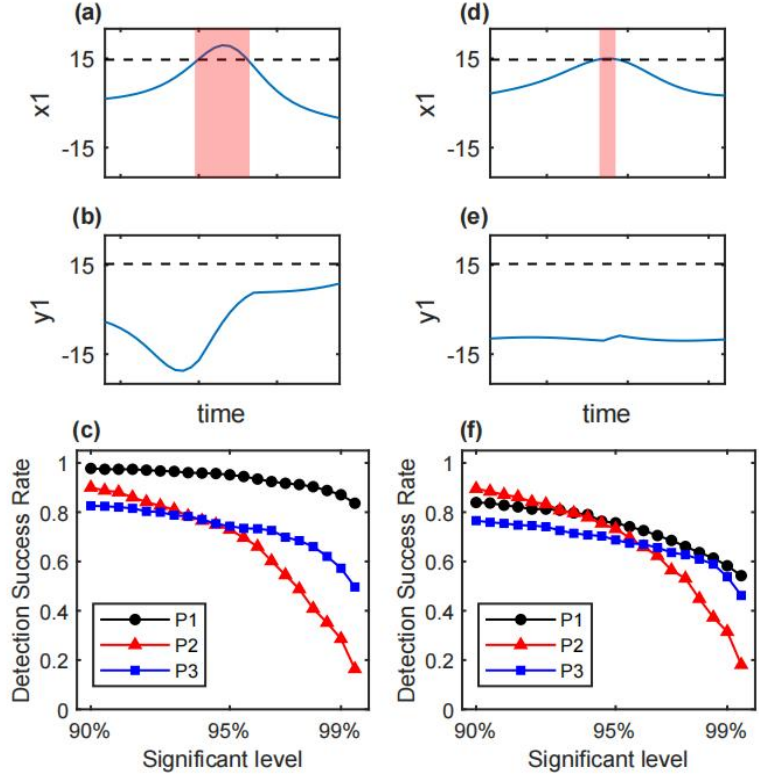

**Fig. 13** Robustness test on intensities of extreme events in causality detecting width $dt = 11$ in the coupled Lorenz-Lorenz model with a unidirectional extreme event-driven coupling. (a), (b), (c) high-intensity case with a maximum value exceeding the 98th percentile, and (d), (e), (f) low-intensity case with a maximum value exceeding the 95th percentile but below the 98th percentile.

Lastly, the impact of co-occurrence of extreme events on detecting event-to-event causality by E2ECD is checked. Among all extreme event drivers, 23% have their own extreme event response, but 77% do not have their counterparts. Event-to-event causality detection by E2ECD was compared between two cases: one with co-occurring driving and driven extreme events, and one with only driving events. The results show that co-occurrence improves both the efficiency and accuracy of detection across all three strategies (P1, P2 and P3, **Fig. 14**). Among these three ways, both P1 (full measurements) and P3 (partial measurements together with state space reconstruction) increase more significantly, at higher significance level such as 98%, the detection success rate remains higher than 70% for P3 and 90% for P1. As global warming increases, more and frequent concurrent extreme events occur [4-9, 16-18]. Most studies focus on the links between concurrent extreme events, but only few devote to cause-effect relationships between them, especially to the causality relationship between two specific extreme events. Success of E2ECD method shown in **Fig. 14** indicates that many related studies can be completed with applying E2ECD.

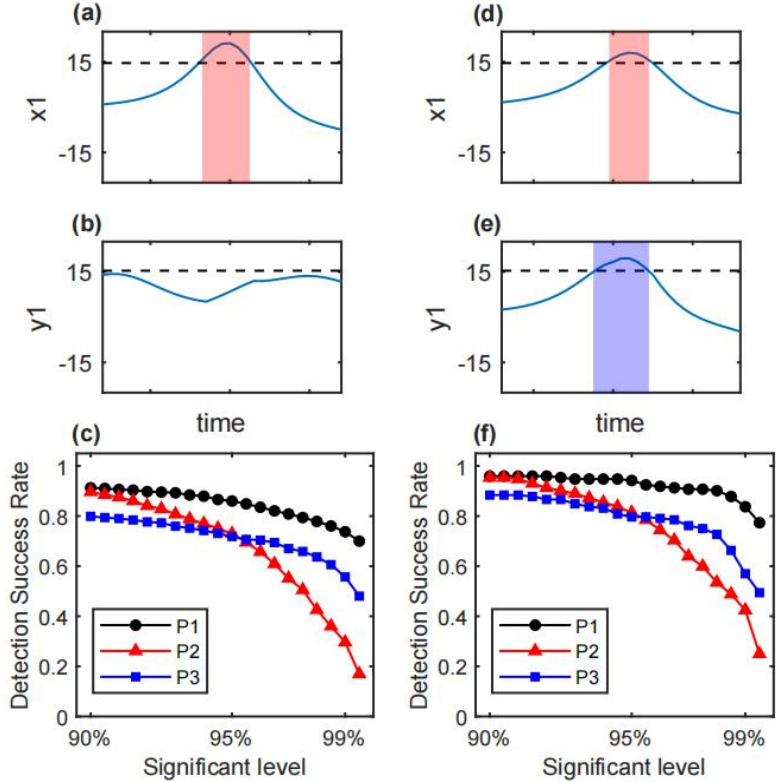

**Fig. 14** Robustness test on co-occurrence of extreme events in causality detecting width $dt = 11$ in the coupled Lorenz-Lorenz model with a unidirectional extreme event-driven coupling. (a), (b), (c) for extremes not occurring simultaneously, and (d), (e), (f) for simultaneous extreme events.

### 4.2 Event-to-event causality detection in Walker circulation system

Walker circulation is a large-scale atmospheric circulation pattern in the tropical Pacific driven by the temperature gradient between western and eastern Pacific [46]. Previous studies have found that states above the mean exhibit stronger causal influence compared to those below [47] and causal interactions between sea surface temperatures and surface temperature anomalies emerge only when these variables deviate coherently from their mean values [37]. To detect and quantify event-to-event causality between these extreme events, the causal index is directly calculated either using the anomalies of the regionally averaged two variables (WPACa and CEPACa), or using WPACa and CEPACa together with the reconstructed state space (with embedding dimension m = 3 and time delay lag = 1). All anomalies in this study are computed by removing daily climatology defined over the period from 1980 to 2021.

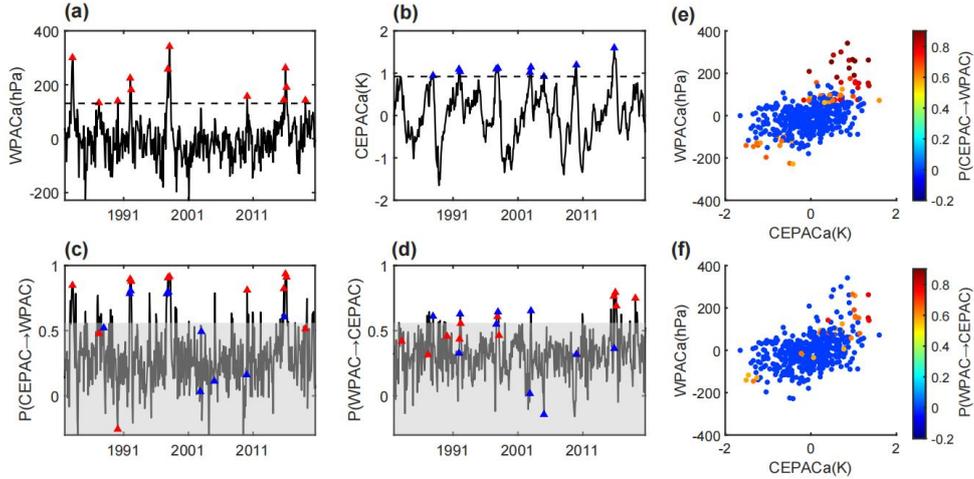

**Fig. 15** Causality detection between WPAC and CEPAC. Time series of WPAC (a) and CEPAC (b). The dashed line represents the 95th percentile, and the triangles represent the peak positions exceeding the 95th percentile. Causality index (c) from CEPAC to WPAC and (d) from WPAC to CEPAC calculated using the regional averaged series. The grey shading represents the 95% significance interval. Scatter plot for WPAC v.s. CEPAC color-shading with the causality index (e) from CEPAC to WPAC (f) from WPAC to CEPAC (values not exceeding the 95% significance interval have been set to 0 with blue shading).

First of all, there are several dominant extreme events in both WPACa and CEPACa series marked by red and blue triangles (**Fig. 15a** and **15b**), respectively. These extreme events interact with one another through the connection of Walker circulation. Secondly, further causality analysis directly from the region-averaged anomaly series shows that causal interactions between these extreme events are asymmetrical, with stronger causal influence from CEPAC to WPAC (**Fig. 15c**) than from WPAC to CEPAC (**Fig. 15d**), which is consistent with the observation that positive CEPAC temperature anomalies drive the positive WPAC surface pressure, not vice versa (**Fig. 15e** and **15f**). More detailed results are summarized in **Table 1**, in which more fraction of asymmetric causal extreme drives can be captured at the higher significance level by the E2ECD method, especially for the cases of co-occurring extreme events.

| Significance level | WPAC+ | | CEPAC+ | | WPAC+/CEPAC+ | |
|---|---|---|---|---|---|---|
| | Pxy | Pyx | Pxy | Pyx | Pxy | Pyx |
| 0.98 | 5/5 | 0/5 | 2/4 | 0/4 | 1/1 | 0/1 |
| 0.95 | 9/12 | 2/12 | 5/10 | 3/10 | 5/5 | 1/5 |
| 0.9 | 15/28 | 4/28 | 10/24 | 8/24 | 7/10 | 4/10 |

**Table 1** Fraction of captured extreme events by causality index calculated using the regional averaged series to total extreme events defined under different significance levels. Pxy for causality from CEPAC to WPAC. Pyx for causality from WPAC to CEPAC. Only WPAC takes its extremes (left), Only CEPAC takes its extremes (middle), both WPAC and CEPAC take their extremes (right).

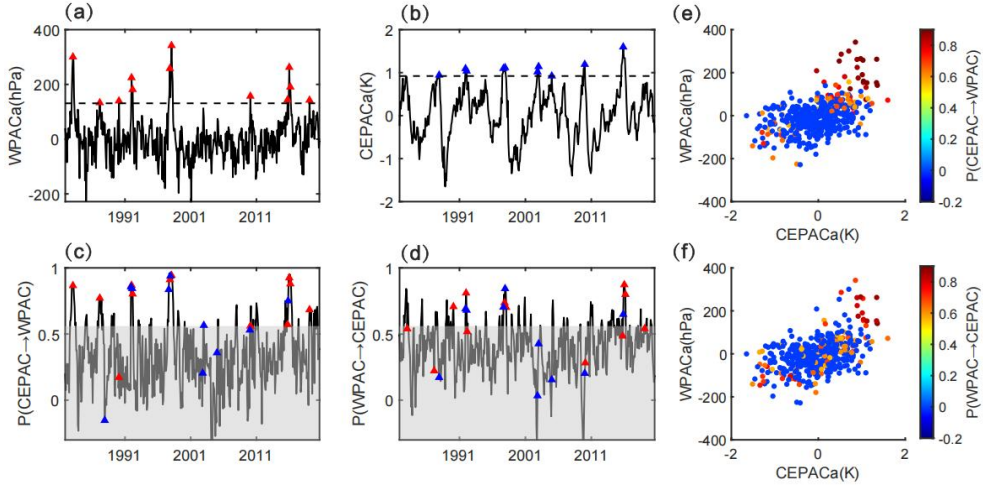

**Fig. 16** Similar as **Fig. 15**, but causality index calculated using the regional averaged series together with state space reconstruction with m = 3 and lag = 1.

When state space reconstruction is additionally involved, similar results can be revealed (**Fig. 16**, **Table 2**). Meanwhile, higher causality can be detected from WPAC to CEPAC (**Fig. 16d 16f**) compared to those without embedding, which may be caused by the loss of instantaneous cause-effect relations with embedding. For the co-occurrence of extreme events, embedding can help E2ECD capture the extreme drivers more precisely (**Table 2**), even at the 90% significance level, the fraction of captured extreme events by causality index is also 100% (**Table 2**).

| Significance level | WPAC+ | | CEPAC+ | | WPAC+/CEPAC+ | |
|---|---|---|---|---|---|---|
| | Pxy | Pyx | Pxy | Pyx | Pxy | Pyx |
| 0.98 | 5/5 | 1/5 | 4/4 | 0/4 | 1/1 | 0/1 |
| 0.95 | 11/12 | 2/12 | 6/10 | 3/10 | 5/5 | 1/5 |
| 0.9 | 22/28 | 5/28 | 13/24 | 7/24 | 10/10 | 1/10 |

**Table 2** Similar as Table 1 but causality index calculated using the regional averaged series together with state space reconstruction with m = 3 and lag = 1.

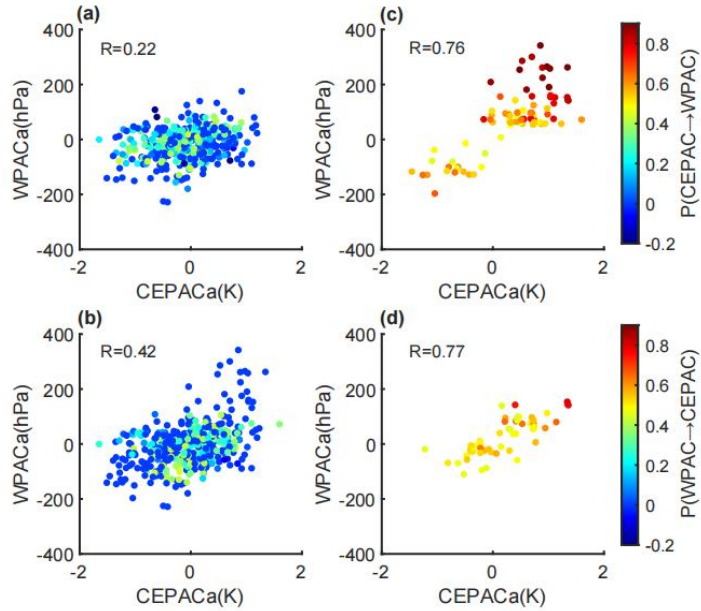

**Fig. 17** Scatter plots of WPACa v.s. CEPACa conditioned on causality index. (a) and (b) for causality indices less than 0.45 corresponding their 90$^{th}$ threshold, (c) and (d) for causality indices larger than 0.45. The values of R show the corresponding correlation coefficient.

To the end, analyzing interactions between surface subsystems in Walker circulation clearly reveals the differences and similarity between correlation and causality (**Fig. 17**), Scatter plots between CEPAC temperature anomalies and WPAC surface pressure anomalies are conditioned on causality index. As expected, extreme events in positive CEPAC temperature anomalies more easily drive their counterparts in WPAC surface pressure anomalies, vice versa (**Fig. 17c** and **17d**) with higher causality indices together with higher correlation between the CEPAC temperature anomalies and the WPAC surface pressure anomalies. However, not all extreme events in positive or negative CEPAC temperature anomalies drive their counterparts in WPAC surface pressure anomalies, vice versa (**Fig. 17a** and **17b**). Lower causality indices are along with lower correlation between the CEPAC temperature anomalies and the WPAC surface pressure anomalies. Especially, most of extreme events in positive WPAC surface pressure anomalies don't drive extreme events in positive CEPAC temperature anomalies, although these extreme events in positive WPAC surface pressure anomalies lead to increase of correlation between CEPAC temperature anomalies and WPAC surface pressure anomalies (**Fig. 17b**). These results further confirm the existence of asymmetrical causal interactions between these extreme events with stronger causal influence from CEPAC to WPAC (**Fig. 15c**) than from WPAC to CEPAC (**Fig. 15d**). These novel findings are not reported in the previous literature.

## 5. Conclusion and discussion

In this study, to answer the question "whether a specific extreme event in a process related to one variable is a cause or a response of another specific extreme event in a process related to another variable", a novel event-to-event causality detecting (E2ECD) method is developed to detect causality between events, with superior performance in detecting causality between co-occurring extreme events. To validate the causality detection performance of newly developed method, the coupled Lorenz-Lorenz model is modified to be driven only by extreme events, which mimics the situations often happening in the real worlds, such as linkage between Pakistan flooding and East Asian heatwaves during July to August 2010 and 2022 [7, 8], but not Pakistan rainfalls linked with East Asian air temperature variations. The applications of traditional causality detecting approaches to this modified coupled model show the limitation of global or local mean causality detecting approaches as they are all designed to provide mean causality detection, which fail to detect even-to-event causality. However, our proposed E2ECD method can indeed deal with this kind of problems, and robust to the hyper-parameters and experiment settings. The successful applications of this method to study the interactions between surface subsystems of Walker circulation indicate the possible applicability in broad real-world scenarios, such as assessing impacts from extreme events [1-3, 11], causality detection between spatial or temporal concurrent compound events [4-9], and so on.

Additionally, our analysis suggests that that the perfect performance of the proposed E2ECD method can only be achieved with full measurements of the studied systems. For the real-world systems, only partial measurements are available. Although integrating state space reconstruction can improve the detecting ability, suitable parameters must be optimized. On the other hand, regional averaging time series such as the index of Walker circulation [37, 39, 40] loses the underlying spatial interactions among spatial units [18, 37, 39, 40]. How to apply these missed spatial interactions in event-to-event causality detecting is an open question, which deserves deep study in future work.

**Declaration of Competing Interest**
The authors declare that they have no known competing financial interests or personal relationships that could have appeared to influence the work reported in this paper.

**Credit authorship contribution statement**
**S.Y. Yu:** Investigation, Software, Validation, Visualization, Writing-original draft. **Y. Huang:** Validation, Writing-review & editing. **Z.T. Fu:** Conceptualization, Supervision, Funding acquisition, Writing-review & editing.

**Acknowledgements**
This work was supported by the National Natural Science Foundation of China (Grant Nos. 42175065 and 42475055).

**Data availability statement**

All data that support the findings of this study are included within the article.

**ORCID IDs**
Yu Huang   https://orcid.org/0000-0002-7930-9056
Zuntao Fu   https://orcid.org/0000-0001-9256-8514